\documentclass[9pt,twocolumn,twoside]{opticajnl}
\journal{opticajournal} % use for journal or Optica Open submissions

\setboolean{shortarticle}{true}
\usepackage{amsmath,amssymb,mathrsfs}
\usepackage{cases}
\usepackage{graphicx}% Include figure files
\usepackage{dcolumn}% Align table columns on decimal point
\usepackage{bm}% bold math
\usepackage{color, xcolor}
\usepackage{amsfonts}
\usepackage{enumerate}
\usepackage{rotating}
\usepackage{epstopdf}
\usepackage{braket}
\usepackage{bbding,pifont}

\definecolor{mygreen}{RGB}{0, 128, 0}

\hypersetup{colorlinks=true, citecolor=blue, linkcolor=blue, urlcolor=blue}

\usepackage{hyperref}
\hypersetup{
	colorlinks=true,
	linkcolor=blue,
	filecolor=blue,
	urlcolor=blue,
	citecolor=magenta
}

\newcommand{\Eq}[1]{Eq.~(\ref{#1})}

\newcommand{\Fig}[1]{Fig.~\ref{#1}}

\newcommand{\beq}{\begin{equation}}
	\newcommand{\eeq}{\end{equation}}
\newcommand{\beqa}{\begin{eqnarray}}
	\newcommand{\eeqa}{\end{eqnarray}}
\newcommand{\Beqa}{\begin{eqnarray*}}
	\newcommand{\Eeqa}{\end{eqnarray*}}

\def\bal#1\eal{\begin{align}#1\end{align}}
\def\Bal#1\Eal{\begin{align*}#1\end{align*}}

\newcommand{\dif}{\mathrm{d}}

\newcommand{\mre}{\mathrm{Re}}	%Real part
\newcommand{\vect}[1]{\mathbf{#1}}

\usepackage{lineno}
\title{Polarization-selective quantum cooperative response in dual-species atom arrays} 

\author[1,2,$\dagger$]{Huan Wang} 
\author[1,2,3,$\dagger$]{Shangguo Zhu} 
\author[1,2,3]{Yun Long} 
\author[1,2,3]{Fei Zhang} 
\author[1,2,3]{Yinghui Guo} 
\author[1,2,3]{Mingbo Pu} 
\author[1,3,*]{Xiangang Luo} 

\affil[1]{State Key Laboratory of Optical Field Manipulation Science and Technology, Institute of Optics and Electronics, Chinese Academy of Sciences, Chengdu 610209, China} 
\affil[2]{Research Center on Vector Optical Fields, Institute of Optics and Electronics, Chinese Academy of Sciences, Chengdu 610209, China} 
\affil[3]{College of Materials Science and Opto-Electronic Technology, University of Chinese Academy of Sciences, Beijing 100049, China} 

\affil[$\dagger$]{These authors contributed equally to this work.} 
\affil[*]{lxg@ioe.ac.cn}

\begin{abstract} 

Atom arrays have emerged as a powerful platform for quantum light-matter interfaces, yet single-species arrays are constrained by
in-plane symmetry, restricting polarization control. 
Here we investigate the cooperative optical response of dual-species subwavelength atom arrays, in which intrinsic polarizability difference
breaks in-plane symmetry. 
By engineering the lattice constants and detunings, the arrays exhibit polarization-dependent subradiant modes, enabling complete reflection of a specific polarization component.  
Leveraging this mechanism, we assemble array units as functional pixels and demonstrate a scalable polarization-selective quantum light modulator. 
Our work establishes a dynamically reconfigurable atomic-photonic platform for versatile subwavelength quantum optical elements. 

\end{abstract}

\setboolean{displaycopyright}{false} % Do not include copyright or licensing information in submission.

\begin{document}
\maketitle

\section{Introduction} 

Atom arrays play a central role in quantum science and technology, offering a highly controllable and versatile platform for frontier areas including quantum computation, quantum simulation, quantum metrology, and quantum communication~\cite{Henriet2020,Browaeys2020,Covey2023,Shen2025}.
In particular, ordered atom arrays with subwavelength spacing provide unique advantages in quantum optics and quantum information processing by enabling strong light-atom interactions and cooperative optical responses~\cite{Reitz2022}.
Recent studies have been focusing on cooperative quantum phenomena such as superradiance and subradiance~\cite{Ballantine2021a,Ballantine2022,Masson2022,Masson2024}, highlighting the potential of atom arrays for quantum information processing, photon storage, and engineered quantum optical elements~\cite{Bekenstein2020,Ballantine2021a,Ballantine2022,Reitz2022,Parmee2023}.

Theoretically, two-dimensional subwavelength atom arrays support cooperative scattering resonances at specific magic lattice constants, enabling near-unity extinction of incident light~\cite{Bettles2016,Shahmoon2017,Javanainen2019,Ruostekoski2023}.
By tailoring array geometries and multilevel atomic configurations, a broad range of optical functionalities have been proposed, including Huygens surfaces, negative refraction, and linear optical elements~\cite{Ballantine2020,Ballantine2021b,Bassler2023,Ruks2025}.
Experimentally, two-dimensional arrays containing a few hundred atoms have been realized as highly efficient optical mirrors~\cite{Rui2020}, and further extended to quantum metasurfaces with dynamic control enabled by auxiliary Rydberg atoms and electromagnetically induced transparency~\cite{Bekenstein2020,Srakaew2023}. 
While these single-species atom arrays offer foundational insights, their inherent symmetry restricts selective control over polarization, suggesting a need for more complex, multi-species configurations.

Recently, dual-species atom arrays have shown distinct advantages in quantum computing, where different atomic species can act as data and auxiliary qubits, enabling parallel quantum operations and fully connected networks~\cite{Sheng2022,Anand2024,Byun2025}. 
More importantly, intrinsic differences in atomic polarizability introduce additional tunable degrees of freedom, opening new possibilities for multi-degree-of-freedom optical field manipulation.

In this work, we systematically study the cooperative optical response of dual-species subwavelength atom arrays and identify polarization-dependent subradiant modes accompanied by pronounced optical anisotropy. 
We reveal the physical origin of polarization selectivity near a symmetric detuning point, which arises from strong coherent scattering with balanced radiative losses.
Leveraging this mechanism, we design and demonstrate a scalable polarization-selective quantum light modulator composed of multiple dual-species array pixels, and discuss feasible experimental implementations.
By bridging fundamental physical insights and device-level architectures, our work opens new avenues for realizing multifunctional quantum light–matter interfaces.

\section{Dual-species atom array}

\begin{figure}[t!]
	\centering
	\includegraphics[width=0.95\linewidth]{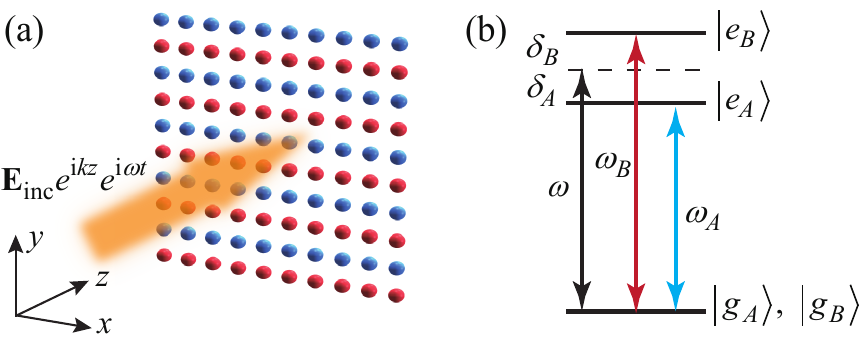}
	%\fbox{\includegraphics[width=0.95\linewidth]{fig_dual-speciesArrayScheme}}
	\caption{(a) Array of two isotopic atomic species, with blue and red spheres denoting atoms of type A and B, respectively. (b) Corresponding two-level energy structure. }
	\label{fig:lattice}
\end{figure}

We consider the cooperative optical response of a two-dimensional subwavelength square array of lattice constant $a$ containing two atomic species arranged in a stripe pattern [Fig.~\ref{fig:lattice}(a)]. 
Each atom is modeled as a two-level system with transition frequency $\omega_A$ or $\omega_B$. 
A light beam incident along the $z$-axis illuminates the array in the $x$–$y$ plane at $z=0$. 
Both species are driven by light of frequency $\omega$, with detunings $\delta_A=\omega-\omega_A$ and $\delta_B=\omega-\omega_B$, as shown in Fig.~\ref{fig:lattice}(b). 
We choose the two species to be isotopes so that they share similar energy levels and can be driven near resonance by the same field, with $\omega$ close to both $\omega_A$ and $\omega_B$.

The incident light is scattered by the atom array with each atom acting as a quantum emitter, which can be modeled as pointlike particles with linear isotropic polarizabilities~\cite{Bettles2016,Shahmoon2017}.  
We consider the cooperative response by analyzing the total electric field, which is the sum of the incident and scattered fields and can be obtained by solving the wave  equation, $\nabla \times \nabla \times \mathbf{E} - k^2 \mathbf{E} = k^2 \mathbf{P} / \varepsilon_0$, where $\mathbf{P}$ is the total polarization, $\varepsilon_0$ is the vacuum permittivity, $k = \omega/c$ is the wave number and $c$ is the light speed. 
The general solution can be expressed as
\begin{align}
	E_{\nu}(\mathbf{r})=E_{\mathrm{inc},\nu}(\mathbf{r})+\frac{k^2}{\varepsilon_0}\sum_{\mu} \int \dif \vect{r}' G_{\nu \mu}(k,\mathbf{r},\mathbf{r}^{\prime})P_{\mu}(\mathbf{r}^{\prime}), \label{Ew1}
\end{align}
where $\nu,\mu = x, y, z$ denote the spatial components. 
$\mathbf{E}_{\mathrm{inc}}(\mathbf{r})$ represents the incident field.	
$G_{\nu\mu}(k,\mathbf{r},\mathbf{r}^\prime)$ is the component of dyadic Green's function, describing the fields produced by an infinitesimal electric current source (see Supplement 1 for explicit expression).

Considering the linear response regime, $\mathbf{P}$ is proportional to the electric field. 
For the discrete atom array, $\mathbf{P}(\mathbf{r})=\sum_{n } \alpha_n\mathbf{E}(\mathbf{r}) \delta(\mathbf{r}-\mathbf{r}_{n})$, where $n$, $\mathbf{r}_{n}$, and $\alpha_n$ denote the index, position, and polarizability, respectively. 
Substituting into \Eq{Ew1} gives the electric field
\begin{align}
	E_{\nu }(\mathbf{r}) &=E_{\mathrm{inc},\nu }(\mathbf{r})+ \frac{4\pi^{2} }{\varepsilon _{0}\lambda ^{2}} \underset{\mu}{\sum }%
	\sum_{n} \alpha_{n} G_{\nu \mu}(k,\mathbf{r},\mathbf{r}%
	_{n})E_{\mu}(\mathbf{r}_{n}),   \label{Ew2}
\end{align}
where $\lambda = 2 \pi/k$ is the wavelength. 
For a closed two-level $J=0$ to $J=1$ atomic
%~\cite{Lambropoulos2007,Craig2012} 
transition~\cite{Lambropoulos2007}, the polarizability takes the form $\alpha_n = -\frac{3 \varepsilon _{0}\lambda^{3} }{4\pi ^{2}}\frac{\gamma/2}{ \delta_n+\mathrm{i}\gamma/2 }$ for small detuning. 
For atoms of type A or B, the subscript $n$ is replaced by A or B.
The spontaneous emission rate $\gamma$ is taken to be the same for both isotopes (see Supplement 1).
Equation~\eqref{Ew2} shows that the local field is the sum of the incident field and the fields scattered by all other emitters. 
Evaluated at the atomic positions, it becomes a self-consistent set of equations that can be solved numerically for arbitrary incident fields (see Supplement 1).

To quantify the scattering of the dual-species array, we define the transmission coefficient by calculating the total power collected over a two-dimensional circular area located at $z = z_L = 150 \lambda $, centered at $x = y = 0$ in the $x$-$y$ plane, with radius $R_L = 90 \lambda$, which is sufficiently large to avoid finite-size effects~\cite{Bettles2016}. 
The collected power is $P = (\varepsilon_0 c^2/2) \int_\mathcal{L}\mre(\mathbf{E}\times\mathbf{B}^\ast)\cdot \dif \vect{A}$, where $\mathcal{L}$ is the circular area and $\dif \mathbf{A} = \dif A  \mathbf{e}_z$. 
For a plane wave, the magnetic field is $\mathbf{B} = \mathbf{e}_z \times \mathbf{E} /c$.
In the absence of atom array, the collected power is $P_{\mathrm{inc}}$. 
Then, the transmission is defined as $T = P/P_{\mathrm{inc}}$. 
Because the stripe pattern breaks the symmetry between the $x$ and $y$ directions, the transmission generally differs for horizontal ($x$) and vertical ($y$) polarizations. 
This polarization dependence provides an additional degree of freedom for manipulating subradiant effects, enhancing the functionality of cooperative atom arrays as quantum light–matter interfaces.

\section{Cooperative optical response}

\begin{figure}[t!]
	\centering
	\includegraphics[width=0.95\linewidth]{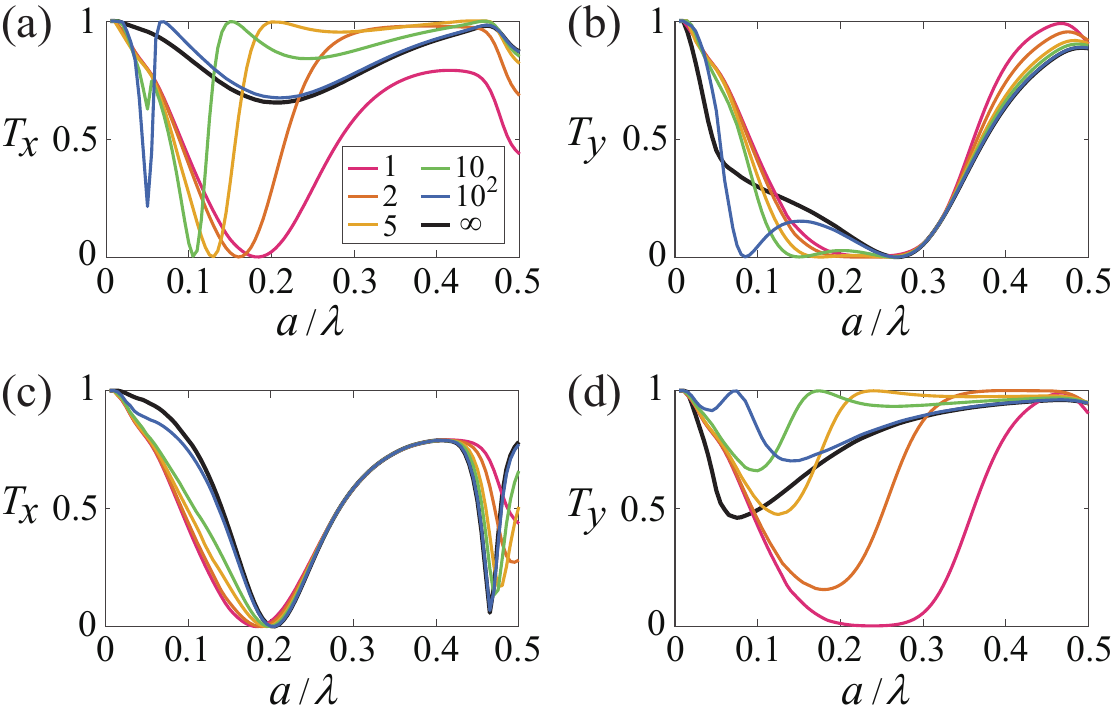}
	\caption{
		Transmissions $T_x$ and $T_y$ as functions of lattice constant $a$. 
		(a,b) The detuning is fixed at $\delta_A/\gamma = 1$, while $-\delta_B/\gamma=1, 2, 5, 10, 100, \infty$. 
		(c,d) The detuning is fixed at $\delta_B/\gamma = -1$, while $\delta_A/\gamma=1, 2, 5, 10, 100, \infty$.
		Different curve colors denote different values of $\delta_A$ or $|\delta_B|$, as indicated in the legend of (a). 
		Calculations assume an $N = 26 \times 26$ atom array and a diagonally polarized incident Gaussian beam with waist $w_0 = 0.3 \sqrt{N} a$. 
	}
	\label{fig:TxTy_varyingdeltaAB}
\end{figure}

In dual-species atom arrays, two detunings $\delta_A$ and $\delta_B$ can be independently tuned to control the atomic polarizabilities and, together with the lattice constant, determine the cooperative optical response. 
We first consider the case where the incident light is near resonant with one atomic species. 
Specifically, we fix $\delta_A$ and vary $\delta_B$, and numerically calculate the transmissions $T_x$ and $T_y$ for horizontally ($x$) and vertically ($y$) polarized incident light. 
We define $T_\mu = P_\mu / P_{\mathrm{inc}, \mu}$ with $\mu=x,y$,  where $P_\mu$ and $P_{\mathrm{inc},\mu}$ are the transmitted and incident optical powers in the $\mu$-polarized component, respectively.

The resulting $T_x$ and $T_y$ as functions of $a$ are shown in \Fig{fig:TxTy_varyingdeltaAB} (a,b). 
For sufficiently large $|\delta_B|$, type B atoms become effectively transparent, and the dual-species square array reduces to an effective single-species rectangular array with lattice constants $a_x = a$ and $a_y = 2a$, corresponding to the curves with $|\delta_B| =\infty$ (see Supplement 1 for detailed analysis). 
For smaller $|\delta_B|$, both $T_x$ and $T_y$ exhibit transmission zeros at distinct lattice constants, indicating polarization-dependent subradiant modes. 
At even smaller $|\delta_B|$, $T_y$ shows two transmission zeros, reflecting enhanced collective response when both atomic species are near resonance. 
Overall, a pronounced anisotropy between the two orthogonal polarizations is observed.

A slightly different picture occurs when $\delta_B$ is fixed and $\delta_A$ is varied, as shown in \Fig{fig:TxTy_varyingdeltaAB} (c,d). Here, the transition of type B atoms is red-detuned with $\delta_B/\gamma=-1$. For sufficiently large $\delta_A$, the array again reduces to an effective single-species rectangular array, corresponding to the curves with $\delta_A=\infty$. 
At smaller $\delta_A$, both $T_x$ and $T_y$ exhibit polarization-dependent subradiant modes, also indicating an enhanced collective response when both atomic species are near resonance.

When both detunings $\delta_A$ and $\delta_B$ are not too large, the dual-species array exhibits strong cooperative effects, leading to substantial suppression of transmission (see Supplement 1 for detailed analysis).  
We then consider the case where the light frequency lies between the two transition frequencies $\omega_A$ and $\omega_B$. 
A particularly symmetric choice is the midpoint, $\delta_A=-\delta_B=\delta$. With identical spontaneous emission rate $\gamma$, this leads to $\mathrm{Re}(\alpha_A)=-\mathrm{Re}(\alpha_B)$ and $\mathrm{Im}(\alpha_A)=\mathrm{Im}(\alpha_B)$ (see Supplement 1).
In this configuration, the two atom species have balanced radiative loss, creating an exotic interference environment that enables strong collective effects through coherent scattering. 

\begin{figure}[t!]
	\centering
	\includegraphics[width=0.45\textwidth]{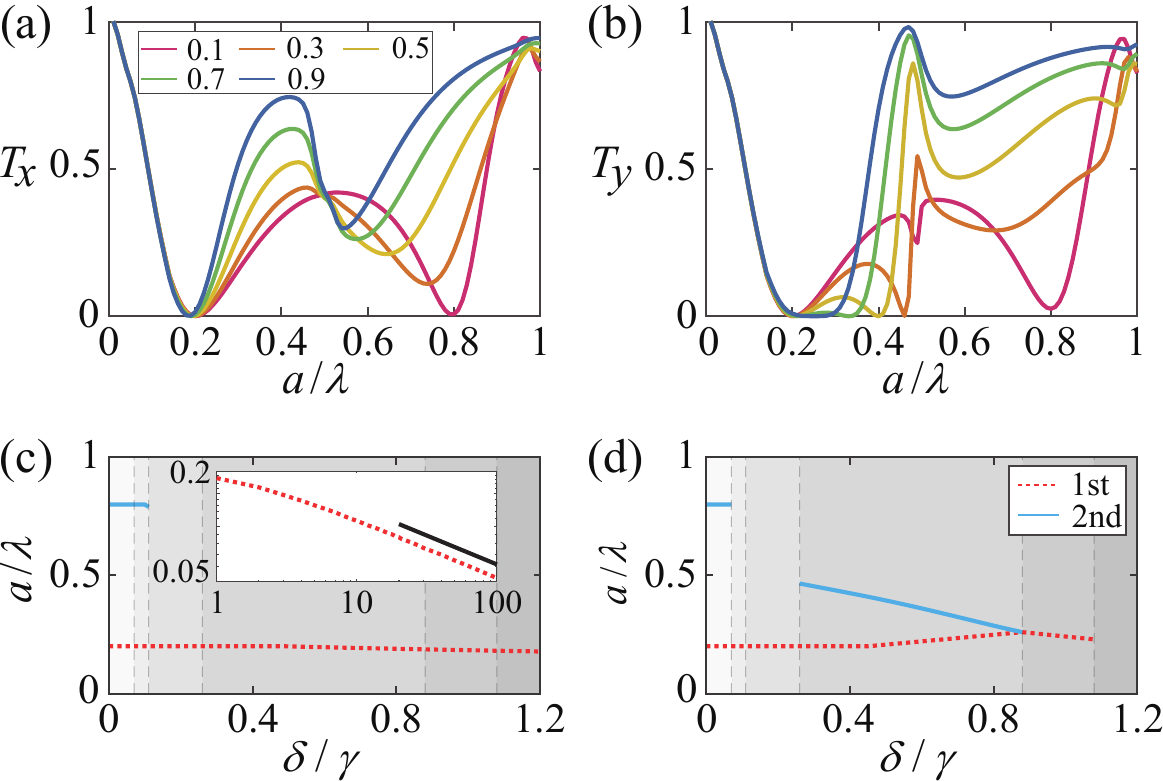}
	\caption{ 
	The transmissions and subradiant modes. 
	(a, b) Transmissions $T_x$ and $T_y$ as functions of the lattice constant $a$. 
	Colors represent $\delta/\gamma = 0.1, 0.3, 0.5, 0.7, 0.9$ (see legend in (a)).
	(c, d) Lattice constants yielding zero transmission versus detuning $\delta$, corresponding to $T_x$ and $T_y$, respectively.
	Blue solid and red dotted lines denote the first and second zeros (legend in (d)); 
	The inset of (c) extends the results to a wider range of $\delta/\gamma$. 
	The black straight line shows a power-law fitting $a \sim \delta^{-0.32}$ at large $\delta$.
    Calculations assume the same atom array and incident beam as in \Fig{fig:TxTy_varyingdeltaAB}. 
	}
	\label{fig:delta_a_TxTy}
\end{figure}

For different values of $\delta$, we numerically calculate the transmissions $T_x$ and $T_y$ as functions of lattice constant $a$, extracted from the electric fields solved via the self-consistent equation [\Eq{Ew2}].
In Fig.~\ref{fig:delta_a_TxTy} (a) and (b), we plot $T_x$ and $T_y$ for various detunings $\delta$. 
At $\delta=0$, the array effectively reduces to a single-species array, where $T_x = T_y$ and zero transmission occurs at $a/\lambda \approx 0.2$, or $0.8$, corresponding to cooperative resonances, consistent with previous single-species square array calculations~\cite{Bettles2016,Shahmoon2017}.

As $\delta$ increases, the dual-species nature emerges, and $T_x$ and $T_y$ diverge as the stripe pattern introduces anisotropy between the $x$- and $y$-directions. 
At larger $\delta$, the number of transmission zeros decreases for both $T_x$ and $T_y$.
The lattice constants for zero transmission as functions of $\delta$ are shown in Fig.~\ref{fig:delta_a_TxTy} (c) and (d). 
In Fig.~\ref{fig:delta_a_TxTy} (c), for small detunings ($0< \delta/\gamma < 0.11 $), $T_x$ has two zeros at $a/\lambda \approx 0.2$ and $0.8$.  
For $\delta/\gamma > 0.11$, one zero around $a/\lambda \approx 0.8$ disappears, leaving only one zero. 
As $\delta/\gamma$ increases, the remaining transmission zero shifts toward smaller $a/\lambda$ and, at large $\delta/\gamma$, approximately follows a numerical power-law fitting, $a/\lambda \sim 0.22 (\delta/\gamma)^{-0.32}$ (see the inset of Fig.~\ref{fig:delta_a_TxTy}(c)).

In Fig.~\ref{fig:delta_a_TxTy} (d), similarly at small detunings ($0<\delta/\gamma < 0.07 $), $T_y$ has two zeros at $a/\lambda \approx 0.2$ and $0.8$. 
As $\delta$ increases from 0, the system evolves from a single-species array to a dual-species array with increasing anisotropy between the $x$- and $y$-directions (see Supplement 1 for detailed analysis). 
One zero disappears at $\delta/\gamma \approx 0.07$, reappears at $\delta/\gamma \approx 0.26$, and merges with the other zero at $\delta/\gamma \approx 0.88$. 
No zero transmission exists for $\delta/\gamma \gtrsim 1.08$.

\begin{figure}[t!]
	\centering
	\includegraphics[width=0.95\linewidth]{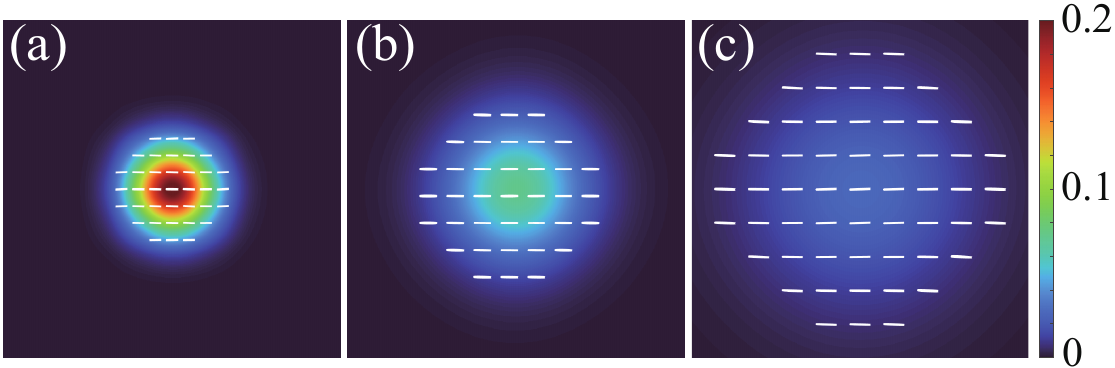}
	\caption{
		(a–c) Total light intensity and polarization distributions at propagation distances $z/\lambda = 14$, $50$, and $90$, respectively. 
		The color bar denotes the normalized optical intensity $I=(|E_x|^2+|E_y|^2)/|E_0|^2$, where $E_0$ is the peak amplitude of the incident Gaussian beam. 
		White lines indicate the polarization distributions. 
		Calculations assume $a/\lambda=0.4$ and $\delta/\gamma=0.5$. 
		The figures span spatial ranges $x/\lambda, y/\lambda \in \left[-10, 10\right]$. 
	}
	\label{fig:IntensityPolar}
\end{figure}

The anisotropy between $T_x$ and $T_y$ provides a novel route to engineer subradiant modes in a polarization-selective manner. 
By choosing appropriate $\delta$ and $a$, one polarization state can enter a subradiant mode with near-complete reflection, while the orthogonal polarization remains transmissive. 
To further illustrate this effect, Fig.~\ref{fig:IntensityPolar} shows the light intensity and polarization distribution for a diagonally polarized incident beam with $\delta/\gamma=0.5$ and $a/\lambda=0.4$. 
The dual-species array nearly completely reflects the $y$-polarized component with $T_y \approx 0$, while producing $x$-polarized light with $T_x \approx 0.50$. 
Remarkably, the transmissive polarization is parallel to the stripe-pattern direction. 
This demonstrates dual-species atom arrays acting as subwavelength optical polarizers, enabling programmable and independent polarization control at the atomic scale.

\section{Polarization-selective quantum light modulator}  

By introducing an ancillary atom which can be excited to a Rydberg state, the atom array can be switched between fully reflective and transparent regimes via electromagnetically induced transparency and Rydberg blockade~\cite{Bekenstein2020,Srakaew2023}. 
The array thus functions as a quantum metasurface that can exist in a superposition of states reflecting and transmitting the incident photons~\cite{Bekenstein2020,Srakaew2023}. 
In a dual-species atom array, such a quantum metasurface becomes polarization-selective, enabling control over specific polarization states and introducing additional tunable degrees of freedom. 
Multiple quantum metasurfaces can be further assembled into a quantum light modulator, with each quantum metasurface acting as a pixel, allowing the generation of highly entangled photonic states~\cite{Bekenstein2020}.

\begin{figure}[t!]
	\centering
	\includegraphics[width=0.95\linewidth]{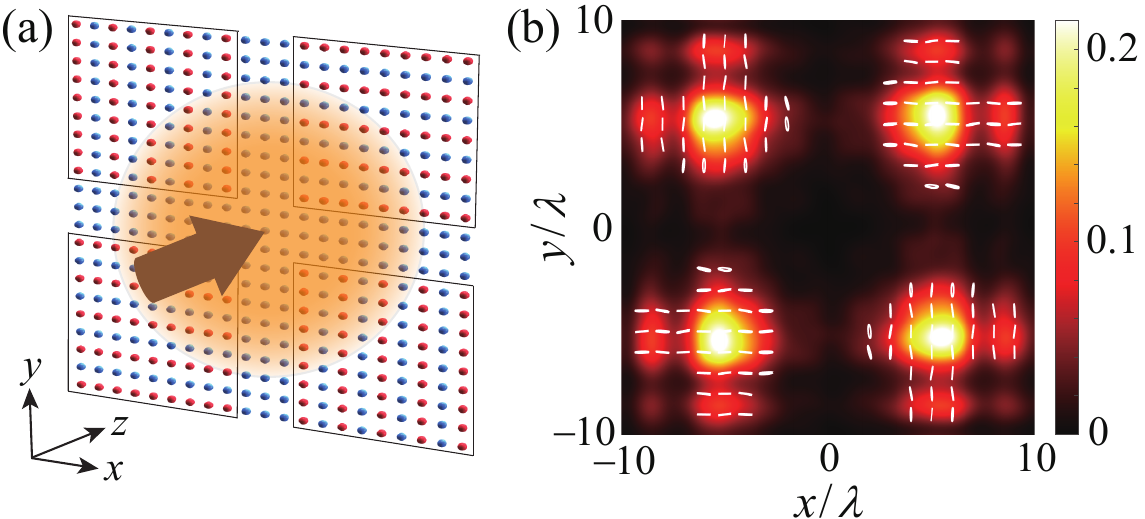}
	\caption{
	(a) Patterned superarray with $2\times 2$ pixels, where each pixel is a dual-species stripe-pattern atom array. 
	(b) Total light intensity and polarization distribution at $z/\lambda = 14$. 
	Calculations assume an $N = 71 \times 71$ atom array with $a/\lambda=0.4$, and a diagonally polarized incident beam with waist $w_0 = 0.3 \sqrt{N} a$ and $\delta/\gamma=0.5$. 
	Each pixel contains $30 \times 30$ atoms and the isolation width is $11$ lattice sites. 
	The color bar denotes the normalized optical intensity $I=(|E_x|^2+|E_y|^2)/|E_0|^2$. 
	White lines or ellipses indicate the polarization distribution.
	}
	\label{fig:assemble}
\end{figure}

Building on this idea, dual-species atom arrays can be used to construct a polarization-selective quantum light modulator, where each pixel selects a desired polarization state. 
We demonstrate this concept by assembling multiple dual-species atom arrays as pixels, forming a programmable platform for pixel-by-pixel polarization control. 
Figure~\ref{fig:assemble} (a) shows a superarray composed of $2\times 2$ pixels, where each pixel is a dual-species atom array with a prescribed stripe-pattern direction (see Supplement 1 for additional configurations). 
The lattice sites between pixels are filled with type-A atoms, serving as isolation regions. 
At $\delta_A/\gamma = 0.5$ and lattice constant $a/\lambda = 0.4$, a square array of type-A atoms exhibits near-perfect reflection for both polarizations with $T_x = T_y \approx 0$, thereby suppressing crosstalk between neighboring pixels.

We further consider a superarray arranged similarly to Fig.~\ref{fig:assemble} (a), with a total atom number $N= 71 \times 71$, where each pixel contains $30 \times 30$ atoms and the isolation width is $11$ lattice sites. 
The spatial distributions of light intensity and polarization are shown in Fig.~\ref{fig:assemble} (b), demonstrating the designed pixel-by-pixel modulation of intensity and polarization. 
This establishes a foundation for assembling polarization-selective quantum light modulators~\cite{Bekenstein2020} and highlights dual-species atom arrays as highly tunable quantum photonic systems at the atomic scale, enabling precise manipulation of vectorial optical fields and a quantum light–matter interface beyond the capabilities of conventional subwavelength optical metamaterials~\cite{Guo2021,Luo2022,Wang2024}.

\section{Experimental Proposal} 

Atom arrays of ytterbium (Yb) have emerged as a powerful platform for quantum computation and quantum precision measurement. 
Yb atoms feature a transition from $6s6p ~{}^3P_0$ to $5d6s ~{}^3D_1$ at a telecom-band wavelength $\lambda \approx 1.4 \mu\mathrm{m}$, realizing a $J=0 \rightarrow J=1$ two-level system, enabling confinement in optical lattices or tweezer arrays with subwavelength spacing. 
We consider a dual-species array of the isotopes $^{171}$Yb and $^{173}$Yb. 
Their transition frequencies are $f_{171} = 215 869 124.9 \mathrm{MHz}$ and $f_{173} = 215 869 151.9 \mathrm{MHz}$, respectively~\cite{Zhao2024}.
Using the spontaneous emission formula 
$\gamma={d^{2}\omega^{3}}/(3\pi \varepsilon _{0}\hbar c^{3})$, we obtain $|f_{173}-f_{171}| \approx 27 \mathrm{MHz} \approx 4.55 \gamma$. 
This energy offset can be continuously tuned via transition frequency shifts induced by AC Stark effects by trapping the two species at different trap depths (see Supplement 1 for a detailed proposal), enabling controllable detunings within the range $0 < \delta < 1.2\gamma$ considered in Fig.~\ref{fig:delta_a_TxTy}. 
Moreover, additional calculations show that positional uncertainty of trapped atoms does not compromise experimental observability (see Supplement 1).

\section{Summary}

Our work systematically investigates the cooperative optical response and polarization control mechanisms in dual-species atom arrays. 
By independently tuning the light detunings and lattice constants, we realize polarization-dependent subradiant modes and identify strong collective scattering at a symmetric detuning point with balanced radiative losses. 
Exploiting this mechanism, the array operates as a subwavelength polarizer with atomic-scale polarization selectivity. 
We further propose and validate a scalable polarization-selective quantum light modulator, enabling spatially programmable polarization control through pixel integration and engineered isolation. 
These results establish dual-species atom arrays as a highly tunable quantum photonic platform and provide a theoretical foundation for advanced quantum light–matter interfaces.

\section{Back matter}

\begin{backmatter}
\bmsection{Funding} S.Z. acknowledges support from the Youth Innovation Promotion Association, Chinese Academy of Sciences (2023399). X.L. acknowledges support from the National Natural Science Foundation of China (U24A6010).

% \bmsection{Acknowledgment} Blank..........

\bmsection{Disclosures} The authors declare no conflicts of interest.

\bmsection{Data Availability Statement} Data underlying the results presented in this paper are not publicly available at this time but may be obtained from the authors upon reasonable request.

\bmsection{Supplemental document}
See Supplement 1 for supporting content. 
\end{backmatter}

% Bibliography
\bibliography{references}

% Full bibliography added automatically for Optics Letters submissions; the following line will simply be ignored if submitting to other journals.
% Note that this extra page will not count against page length

\bibliographyfullrefs{references}

% Please include bios and photos of all authors for aop articles
%\ifthenelse{\equal{\journalref}{aop}}{%
%	\section*{Author Biographies}
%	\begingroup
%	\setlength\intextsep{0pt}
%	\begin{minipage}[t][6.3cm][t]{1.0\textwidth} % Adjust height [6.3cm] as required for separation of bio photos.
%		\begin{wrapfigure}{L}{0.25\textwidth}
%			\includegraphics[width=0.25\textwidth]{john_smith.eps}
%		\end{wrapfigure}
%		\noindent
%		{\bfseries John Smith} received his BSc (Mathematics) in 2000 from The University of Maryland. His research interests include lasers and optics.
%	\end{minipage}
%	\begin{minipage}{1.0\textwidth}
%		\begin{wrapfigure}{L}{0.25\textwidth}
%			\includegraphics[width=0.25\textwidth]{alice_smith.eps}
%		\end{wrapfigure}
%		\noindent
%		{\bfseries Alice Smith} also received her BSc (Mathematics) in 2000 from The University of Maryland. Her research interests also include lasers and optics.
%	\end{minipage}
%	\endgroup
%}{}

\end{document}

\maketitle

\section{Dyadic Green's function}
The free-space dyadic Green's function under the resonance condition ($\omega = c k$) is defined as~\cite{Novotny2012} 
\begin{align}
	\overline{\overline{\mathbf{G}}} (\mathbf{r},\mathbf{r}^{\prime})=\left[ \overline{\overline{\mathbf{I}}} + \frac{1}{k^2}\nabla\nabla\right] \frac{\mathrm{e}^{\mathrm{i} k R}}{4 \pi R},
\end{align}
which connects a localized electric dipole source to the radiated field. 
The operator $\nabla\nabla$ denotes the dyadic product, $\overline{\overline{\mathbf{I}}}$ is the unit dyadic, and $\mathbf{R}=\mathbf{r}-\mathbf{r}'$ with $R=|\mathbf{R}|$.

The components of dyadic Green's function are given by
\begin{align}
	G_{\nu \mu }(k,\mathbf{r},\mathbf{r}^\prime)=\frac{e^{\mathrm{i} k R}}{4\pi R}\left[\left(1+\frac{\mathrm{i}kR-1}{k^2R^2}\right)\delta_{\nu \mu } +\left(-1+\frac{3-3\mathrm{i}kR}{k^2 R^2}\right)\frac{R^{\nu} R^{ \mu }}{R^2}\right], \label{Gij}
\end{align}	
with  $R^{\nu}=\mathbf{e}_{\nu}\cdot \mathbf{R}$. $\nu,\mu = x, y, z$ denote the three components in the Cartesian coordinate.
 
\section{Atomic polarizability}
We consider the cooperative optical response of a square array containing two atomic species, labeled by $A$ and $B$.
Each atom undergoes a $J=0\rightarrow J=1$ transition, characterized by transition frequency $\omega_n$ and dipole matrix element $d_n$. 
The subscript $n$ is replaced by $A$ or $B$, if the $n$-th atom is type $A$ or $B$, respectively. 
Within linear response, the polarizability of each species is given by~\cite{Lambropoulos2007,Craig2012}
\begin{align}
	\alpha_n
	=
	\frac{2 d_n^2 \omega_n}{\hbar}
	\frac{1}{\omega_n^2-\omega^2-i\gamma_n\omega},
\end{align}
where $\gamma_n=d_n^2\omega_n^3/(3\pi\varepsilon_0\hbar c^3)$ denotes the spontaneous decay rate and $\omega$ is the driving frequency.

For near-resonant driving with the small detuning $\delta_n = \omega - \omega_n$, satisfying $| \delta_n| \ll \omega_n$, the polarizability reduces to
\begin{align}
	\alpha_n
	=
	-\frac{3}{4\pi^2}\varepsilon_0\lambda^3
	\frac{\gamma_n/2}{\delta_n+i\gamma_n/2}.
\end{align}

In the main text, the detunings $\delta_A$ and $\delta_B$ are treated as independently tunable parameters, enabling control over the cooperative optical response of the two atomic species. 
To analyze the detuning dependence of the polarizability, we show two representative level configurations. 
For convenience, we introduce the constant $C_0=3\varepsilon_0\lambda^3/(4\pi^2)$.

In the limit where species B is far off resonant, while species A remains near resonant with the driving field ($\delta_A = 0$ and $0 \ll |\delta_B| \ll \omega_B$), the polarizability of species B satisfies
\begin{align}
	\mathrm{Re}(\alpha _{B}) \sim -C_0\frac{\gamma _{B}/2}{\delta
		_{B}},  \text{\ \ \ }	\mathrm{Im}(\alpha _{B}) \sim C_0\left( \frac{\gamma _{B}/2}{%
		\delta _{B}}\right) ^{2}.  \label{alphaB}
\end{align}
Both the real and imaginary parts of $\alpha_B$ vanish as $|\delta_B|$ increases, indicating that the contribution of species $B$ becomes negligible.
In this regime, the array can be well approximated as a rectangular lattice composed of a single atomic species.

Alternatively, when the driving field is near resonance with both species and its frequency lies at the midpoint between their transition frequencies ($\delta_A = -\delta_B = \delta$), the polarizabilities read
\begin{align}
	\mathrm{Re}(\alpha _{A}) &=-C_0\frac{\delta \gamma /2}{\delta ^{2}+\left(
		\gamma /2\right) ^{2}},  \text{\ \ \ }	\mathrm{Im}(\alpha _{A})=C_0\frac{\left( \gamma /2\right)
		^{2}}{\delta ^{2}+\left( \gamma /2\right) ^{2}}, \nonumber \\
	\mathrm{Re}(\alpha _{B}) &=C_0\frac{\delta \gamma /2}{\delta ^{2}+\left(
		\gamma /2\right) ^{2}},  \text{\ \ \ }	\mathrm{Im}(\alpha _{B})=C_0\frac{\left( \gamma /2\right)
		^{2}}{\delta ^{2}+\left( \gamma /2\right) ^{2}}.
\end{align}
For two isotopic atomic species with  similar energy levels, the spontaneous emission rates are approximately equal, $\gamma_A = \gamma_B = \gamma$. 
In this case, the polarizabilities of species A and B exhibit identical imaginary parts but opposite real parts, providing a unique condition for engineering polarization-dependent collective modes.

\section{Gaussian Driving Field}
The incident field is taken to be a paraxial Gaussian beam propagating along the $z$ axis,
\begin{align}
	\mathbf{E}_{\mathrm{inc}}(x,y,z)=E_0 \frac{w_0}{w(z)}e^{\mathrm{i} k z}e^{-\mathrm{i}\varphi(z)}e^{-\frac{x^{2}+y^{2}}{w^2(z)}}e^{\mathrm{i} k \frac{x^{2}+y^{2}}{2R(z)}} \mathbf{e}_d, \label{Einc}
\end{align} 
where
\begin{align}
	w(z)=w_0\sqrt{1+\left(\frac{z}{z_R}\right)^2},\quad z_R=\frac{\pi w_0^2}{\lambda},\quad R(z)=z\left[1+\left(\frac{z_R}{z}\right)^2\right],\quad\varphi(z)=\arctan\left(\frac{z}{z_R}\right).
\end{align}
Here $E_0$ is the peak amplitude of the incident
Gaussian beam, $w_0$ is the beam waist,  
$R(z)$ is the beam curvature,
$\varphi(z)$ is the Gouy phase and $z_R$ is the Rayleigh range.
In the main text, we consider diagonally polarized light at normal incidence with unit vector $\mathbf{e}_d = 1/\sqrt{2} (\mathbf{e}_x,\mathbf{e}_y, 0)$, where $\mathbf{e}_x$ and $\mathbf{e}_y$ denote the unit vectors in the $x$- and $y$-directions, respectively.

\section{Numerical Method}
For a finite array of point dipoles, the electromagnetic scattering field at any spatial point (see Eq. (2) in the main text) can be obtained numerically via matrix inversion~\cite{Shahmoon2017}.
Specifically, we label the atoms by indices $n = 1, 2, \dots, N$, with positions $\mathbf{r}_n$.
The local electric field experienced by atom $m$ is then given by:
\begin{align}
	E_{\nu}(\mathbf{r}_m) = E_{\mathrm{inc},\nu}(\mathbf{r}_m) + \frac{4\pi^{2} }{\varepsilon_{0}\lambda^{2}} \sum_{n \neq m} \sum_{\mu } G_{\nu \mu }^{mn} \alpha_n E_{\mu }(\mathbf{r}_n). \label{unified_E}
\end{align}
Here, the exclusion of the self-interaction term ($n=m$) reflects the fact that an atom does not respond to its own radiated field. All self-action effects are already incorporated into the complex atomic polarizability $\alpha_n$, while the dyadic Green’s function accounts solely for photon-mediated electromagnetic interactions between distinct atoms~\cite{Shahmoon2017}. 
Notably, the total local field at each atomic site is given by the superposition of the incident field and the fields radiated by all other dipoles in the array, with induced dipole moment $\mathbf{p}_n = \alpha_n \mathbf{E}(\mathbf{r}_n)$.

To solve Eq.~(\ref{unified_E}) efficiently, we rewrite the coupled equations into matrix form. We introduce the $3N$-dimensional field vector
$\overline{E} = (E_x^1, E_y^1, E_z^1, E_x^2, \ldots, E_z^N)^T$,
together with the incident field vector $\overline{E}_{\mathrm{inc}}$ defined analogously. The electromagnetic coupling between atoms is encoded in the $3N \times 3N$ Green’s function matrix $\overline{\overline{\mathbb{G}}}$, constructed from Eq.~(\ref{Gij}). The atomic polarizabilities are incorporated into the diagonal matrix $\overline{\overline{\mathbb{A}}}$, whose elements take the values $\alpha_A$ or $\alpha_B$ depending on the atomic species.
With these definitions, Eq.~(\ref{unified_E}) can be written tightly in matrix form as
\begin{align}
	\overline{E}=\overline{E}_\mathrm{inc}+C\overline{\overline{\mathbb{G}}}(\overline{\overline{\mathbb{A}}}\overline{E}),
\end{align}
where $C = 4\pi^{2}/(\varepsilon_{0}\lambda^{2})$. The self-consistent local field is obtained by direct matrix inversion,
\begin{align}
	\overline{E} = \left[ \overline{\overline{I}} - C \overline{\overline{\mathbb{G}}} \overline{\overline{\mathbb{A}}} \right]^{-1} \overline{E}_{\mathrm{inc}}. \label{Ew_bar_hetero}
\end{align}
Thus, the total electric field at an arbitrary spatial point $\mathbf{r}$ can be evaluated as the superposition of the incident field and the fields radiated by all induced dipoles  (see Eq.~(2) in the main text).

\section{Detuning Dependence of the Transmissions}
To investigate the influence  of detuning on the cooperative response of the dual-species atom array, we calculate transmissions by fixing the lattice constant and scanning the detuning $\delta/\gamma$. The lattice constants are chosen according to the transmission zeros in Fig.~3(b) of the main text, namely $a/\lambda = 0.20, 0.33, 0.40, 0.46$.

\begin{figure}[h!]
	\centering
	\includegraphics[width=0.75\linewidth]{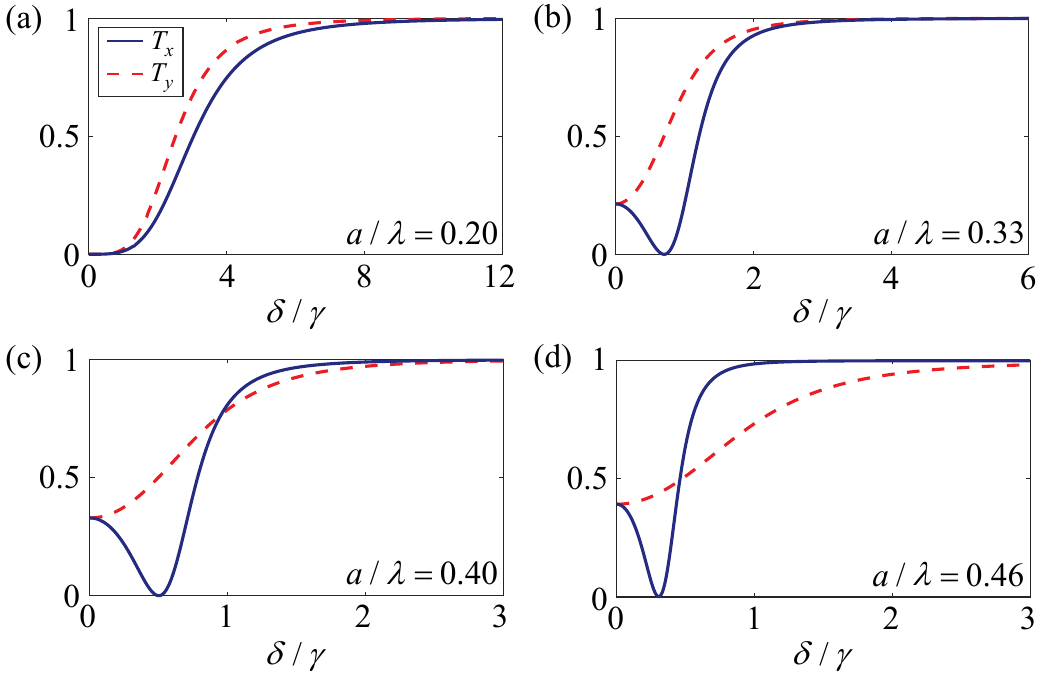}
	\caption{Transmissions $T_x$ and $T_y$ as functions of the detuning $\delta/\gamma$ for different lattice constants: (a) $a/\lambda = 0.20$, (b) $a/\lambda = 0.33$, (c) $a/\lambda = 0.40$, and (d) $a/\lambda = 0.46$.  The transmission zeros of $T_y$ occur at $\delta/\gamma = 0, 0.7, 0.5, 0.3$, respectively, consistent with the results in Fig.3 of the main text.  Calculations assume an $N = 26 \times 26$ atom array and a diagonally polarized incident Gaussian beam with waist $w_0 = 0.3 \sqrt{N} a$. }
	\label{TxTy_delta}
\end{figure} 

The transmissions $T_x$ and $T_y$ are shown in Fig.~\ref{TxTy_delta}(a–d). For all selected lattice constants, both transmissions deviate significantly from unity at small detuning, indicating strong transmission suppression. 
This behavior originates from cooperative scattering in the ordered atomic array. When the detuning is small, the atomic polarizability becomes large, which enhances the induced dipole moments and strengthens the radiative dipole–dipole coupling between atoms. 
As the detuning increases, the atomic polarizability decreases, leading to weaker radiative coupling between atoms. 
Consequently, the collective scattering effect is gradually suppressed, and both $T_x$ and $T_y$ approach unity.
The transmission zeros of $T_y$ appear at $\delta/\gamma = 0, 0.7, 0.5, 0.3$, consistent with the results reported in the main text.

\textbf{Transmission zeros when increasing the detuning from zero.} 
As shown in Fig. 3(c,d) of the main text, as the detuning $\delta$ increases from 0, the transmission zeros changes, indicating qualitative evolving of the cooperative response of the dual-species atom array. 
The critical values $\delta/\gamma = 0.07, 0.26, 0.88, 1.08$ are obtained numerically and mark characteristic detunings where the transmission behavior changes qualitatively. Below we provide a possible qualitative physical interpretation of the underlying cooperative behavior: 

1) Small detuning region ($\delta/\gamma < 0.07$). 
For small detuning, the anisotropy is weak and the optical response closely resembles that of a single-species array. As a result, two transmission zeros appear at $a/\lambda \approx 0.2$ and $a/\lambda \approx 0.8$, consistent with previous studies~\cite{Bettles2016, Shahmoon2017}.

2) Disappearance of a transmission zero ($\delta/\gamma \approx 0.07$).
As the detuning increases, the anisotropy along the $y$-direction modifies the dipole-dipole interference conditions. This breaks the cooperative resonance at the larger lattice spacing ($a/\lambda \approx 0.8$), which is less robust due to weaker cooperative scattering, causing the corresponding zero-transmission point to disappear.

3) Reappearance of a transmission zero ($\delta/\gamma \approx 0.26$).
With further increasing detuning, the geometric phase-matching condition for cooperative scattering can again be satisfied at smaller lattice spacings, leading to the reappearance of a transmission zero. This behavior may be related to geometric resonance and Umklapp scattering mechanisms discussed in Ref.~\cite{Masson2024}.

4) Merging of transmission zeros ($\delta/\gamma \approx 0.88$).
As the detuning increases further, sustaining cooperative effects requires smaller lattice spacings. Consequently, the second transmission zero gradually shifts and eventually merges with the first one.

5) Complete disappearance of transmission zeros ($\delta/\gamma \gtrsim 1.08$).
For sufficiently large detuning, the system moves far from atomic resonance and the atomic polarization response becomes weak, suppressing cooperative scattering. Because the lattice geometry is uniform along the x-direction (resembling a single-species array) but alternating along the y-direction, the zero-transmission condition disappears along y, while one zero-transmission point remains along x (see Fig. 3(c) of the main text).

\section{Effective Polarizability in rectangle array}
In the limit $|\delta_B| \rightarrow \infty$, the polarizability of species B vanishes ($\alpha_B \approx 0$).
Therefore, the array effectively reduces to a rectangular array of species A.
The total electric field then becomes
\begin{align}
	E_{\nu }(\mathbf{r}) =E_{\mathrm{inc},\nu }(\mathbf{r})+\frac{%
		4\pi ^{2} }{\varepsilon _{0}\lambda ^{2}}\underset{\mu }{\sum }%
	\sum_{A} G_{\nu \mu }(k,\mathbf{r},\mathbf{r}%
	_{A}) \alpha_{A} E_{\mu }(\mathbf{r}_{A}).  \label{E_rect}
\end{align}
Here, the primitive vectors are $\mathbf{a}_x = a \mathbf{e}_x = (a, 0, 0)$, $\mathbf{a}_y = 2a \mathbf{e}_y = (0, 2a, 0)$, with a unit cell area $A_0 = 2a^2$. 

To analyze the cooperative optical response, we further consider an infinite rectangular array and expand the incident field in plane-wave components, $ \mathbf{E}_{\mathrm{inc}}(\mathbf{r})=\sum_{\mathbf{k}_{\parallel}}\mathbf{E}_{\mathrm{inc},\mathbf{k}_{\parallel}}e^{\mathrm{i}\mathbf{k}\cdot\mathbf{r}}$ where $\mathbf{k}_{\parallel} = (k_x,k_y)$ is the in-plane wave vector and $k_z = \sqrt{k^2 - |\mathbf{k}_{\parallel}|^2}$.  Applying a Fourier transformation to Eq.~(\ref{E_rect}) for each $\mathbf{k}_{\parallel}$ yields the self-consistent relation
\begin{equation}
	\mathbf{p}(\mathbf{k}_\parallel)=\alpha_A \mathbf{E}_{\mathrm{inc},\mathbf{k}_\parallel}+4\pi^2\frac{\alpha_A}{\varepsilon_0\lambda^3}\lambda\overline{\overline{g}}(\mathbf{k}_\parallel)\mathbf{p}(\mathbf{k}_\parallel), \label{pkll}
\end{equation}
where $\overline{\overline{g}}(\mathbf{k}_{\Vert })= \sum_{A\neq0}\overline{\overline{G}}(0,\mathbf{r}_A)e^{-\mathrm{i} \mathbf{k}_\parallel\cdot\mathbf{r}_A}$, the dyadic Green function $\overline{\overline{G}}(0,\mathbf{r}_A)$ depends only on the relative distance between  central atom ($\mathbf{r}_A = 0$) and other atoms (at $\mathbf{r}_A$).

Solving Eq.~(\ref{pkll}) yields the induced dipole moment in momentum space,  $\mathbf{p}(\mathbf{k}_{\Vert })  = \overline{\overline{\alpha}}_e(\mathbf{k}_\parallel) \mathbf{E}_{\mathrm{inc},\mathbf{k}_\parallel}$,
which defines the effective polarizability tensor
\begin{equation}
	\overline{\overline{\alpha}}_{e}(\mathbf{k}_{\parallel})=-\frac{3}{4\pi^{2}}\varepsilon_{0}\lambda^{3}\frac{\gamma /2}{\delta_A-\overline{\overline{\Delta}}(\mathbf{k}_{\parallel})+i[\gamma+\overline{\overline{\Gamma}}(\mathbf{k}_{\parallel})]/2}, \label{alpha_eff}
\end{equation}
where the cooperative shift and width tensors are given by
\begin{equation}
	\overline{\overline{\Delta}}(\mathbf{k}_\parallel) -\mathrm{i} \frac{\overline{\overline{\Gamma}}(\mathbf{k}_\parallel)}{2}=-\frac{3}{2} \gamma \lambda\sum_{A\neq0}\overline{\overline{G}}(0,\mathbf{r}_A)e^{-\mathrm{i} \mathbf{k}_\parallel\cdot\mathbf{r}_A}.\label{Delta_kGamma_k}
\end{equation}
Equation~(\ref{alpha_eff}) indicates that each in-plane wave vector $\mathbf{k}_{\parallel}$ excites an independent polarization mode of the array, without coupling to other components. Under normal incidence, only the $\mathbf{k}_{\parallel}=0$ mode contributes to the optical response~\cite{Shahmoon2017,Bassler2023}.

\begin{figure}[h!]
	\centering
	\includegraphics[width=0.7\linewidth]{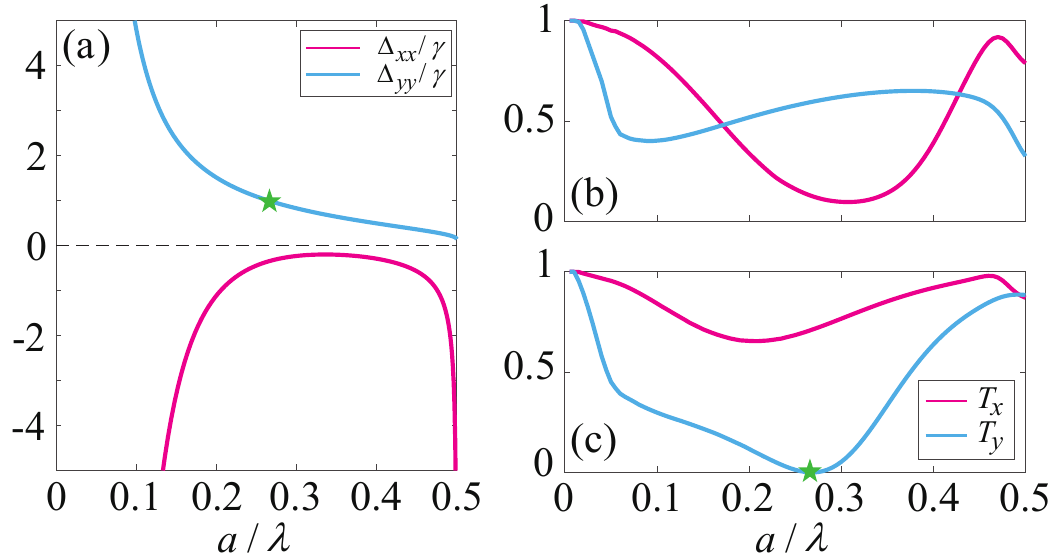}
	\caption{(a) Cooperative shifts as a functions of $a/\lambda$  for a rectangular lattice at normal incidence. (b)(c) Transmissions $T_x$ and $T_y$ as a functions of $a/\lambda$ for $\delta_A/\gamma = 0$, $\delta_A/\gamma = \Delta_{yy}/\gamma = 1$. The detuning of B-type atoms is fixed at $|\delta_B|/\gamma \rightarrow \infty$. The lattice constant denoted by the green star is $a/\lambda = 0.27$. 
	Other parameters are chosen as follows: for panel (a), the number of atoms is $N=10000 \times 10000$; for panel (b)(c), the incident light is diagonally polarized with a beam waist of $w_0 = 0.3 \sqrt{N} a$, the number of atoms is $N = 26 \times 26$. }
	\label{fig:rectangle}
\end{figure}

Using this framework, we compute the cooperative shift tensor $\overline{\overline{\Delta}}$ as a function of the lattice constant $a$, as shown in Fig.~\ref{fig:rectangle}(a). 
In contrast to a square lattice~\cite{Shahmoon2017}, the rectangular geometry breaks in-plane symmetry, leading to a polarization-dependent cooperative response with $\Delta_{xx} \neq \Delta_{yy}$. This anisotropy enables independent tuning of the cooperative resonance conditions for different polarizations.

To confirm this prediction, we also numerically evaluate the transmission of the array under normal incidence. When the A-type atoms are resonant with the incident field ($\delta_A/\gamma = 0$), the transmission remains finite, as shown in Fig.~\ref{fig:rectangle}(b).
This behavior is consistent with the cooperative shifts, since neither $\Delta_{xx}/\gamma$ nor $\Delta_{yy}/\gamma$ crosses zero within the considered parameter range.
By contrast, when the detuning is fixed at $\delta_A/\gamma = 1$, the cooperative condition $\Delta_{yy} / \gamma = 1$  is satisfied at $a/\lambda\simeq0.27$. Consistently, a zero-transmission point appears in the $T_y$ spectrum at the same lattice constant [Fig.~\ref{fig:rectangle}(c)], signaling a cooperative resonance in which the $y$-polarized component of the incident field is completely reflected by the two-dimensional atomic array.

\section{Additional examples of pixel-based polarization-selective modulation}
In this section, we present additional examples that support the pixel-based polarization-selective modulation scheme introduced in the main text. 
The simulations illustrate how different spatial assemblies of the dual-species atomic-array pixel influence the transmitted-field intensity and polarization distributions.

\begin{figure}[h!]
	\centering
	\includegraphics[width=0.9\linewidth]{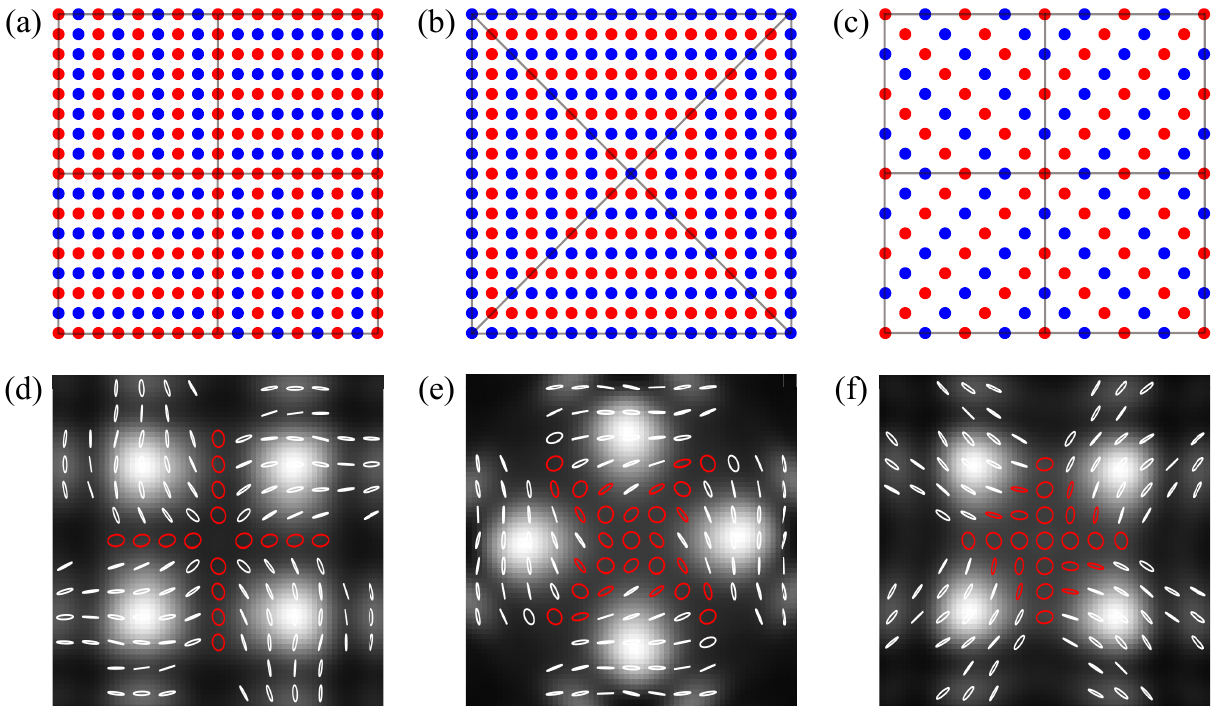}
	\caption{(a–c) Schematics of three composite configurations assembled by dual-species atomic-array pixels. (d–f) Corresponding normalized optical intensity distributions ($I=(|E_x|^2+|E_y|^2)/|E_0|^2$) and local polarization ellipses. The simulations use $N=71 \times 71$ atoms with lattice constant $a/\lambda = 0.4$, illuminated by a normally incident right-handed circularly polarized beam with waist $w_0=0.3\sqrt{N}a$. The atomic detunings are $\delta_A/\gamma = -\delta_B/\gamma =0.5$, and the field is evaluated at a propagation distance $z/\lambda = 14$. The spatial domain is $x/\lambda, y/\lambda \in \left[-10, 10 \right]$, and the color scale spans the range $[0, 0.4]$. White and red represent linear and right-handed elliptical polarization states, respectively.}
	\label{fig:array_Polar}
\end{figure}

We construct three representative 
composite configurations by applying spatial rotations and geometric deformations to the pixel units, 
as illustrated in Figs.~\ref{fig:array_Polar}(a–c). 
The corresponding transmitted-field intensity distributions and local polarization ellipses, evaluated at a propagation distance of $z/\lambda = 14$, are shown in Figs.~\ref{fig:array_Polar}(d–f). 
Although all configurations are illuminated under identical conditions, their near-field optical responses exhibit markedly different polarization patterns. 
In particular, regions of enhanced field intensity correlate strongly with the high-density substructures 
formed by the assembled pixels, indicating that the response is dominated by collective scattering modes 
localized on these sub-arrays.

By comparing with Fig.~4 of the main text, we find that configurations lacking sufficient isolation between neighboring pixel units exhibit weak inter-pixel coupling near their boundaries, which slightly perturbs the designed polarization patterns. 
However, these effects remain moderate for the parameters considered here and do not qualitatively alter the pixel-level polarization selectivity demonstrated in the main text.

Overall, these supplementary examples further support the robustness of the pixel-based polarization-selective modulation mechanism against variations in pixel geometry, providing additional numerical validation of the design principles established in the main text.

\section{Effect of Atomic Positional Uncertainty} 
In realistic experiments, trapped atoms exhibit a finite spatial spread coming from their wave packet of the occupied quantum states, leading to positional uncertainty, i.e., deviations of the atomic positions from the lattice-site centers. 
This uncertainty increases for shallower trapping potentials. 
To quantify the effect of positional uncertainty on the cooperative optical response of a dual-species atomic array, we model the atomic positions by a Gaussian distribution corresponding to the ground state of a harmonic trap~\cite{Bettles2016}.
For a two-dimensional atom array, the axial motion is strongly suppressed, and the $z$ degree of freedom can be integrated out. 
Thus, the in-plane displacement $(X_n,Y_n)$ of the $n$-th atom from its ideal lattice position is described by the probability density
\begin{equation}
	\rho_n(X_n,Y_n)=\frac{1}{\pi l_r^2}\exp\left(-\frac{X_n^2+Y_n^2}{l_r^2}\right),
\end{equation}
where $l_r=a s^{-1/4}/\pi$ characterizes the spatial extent of the atomic wave packet, depending on the normalized trap depth $s=V_0/E_R$, with $V_0$ the trap depth and $E_R=\pi^2\hbar^2/(2ma^2)$ the recoil energy. 

Next, atomic positional uncertainty are incorporated via a Monte Carlo approach~\cite{Bettles2016}: for each realization, displacements are randomly sampled from $\rho_n$ and added to ideal lattice coordinates, and the observation is averaged over many realizations. 
In this part, we mainly focus on a regime corresponding to a Mott-insulator state, where atoms are typically trapped in trap depths $V_0 \sim 20$–$50E_R$, while even deeper traps up to $V_0\sim10^3E_R$ are possible~\cite{Bakr2009, Sherson2010}.
\begin{figure}[h!]
	\centering
	\includegraphics[width=0.75\linewidth]{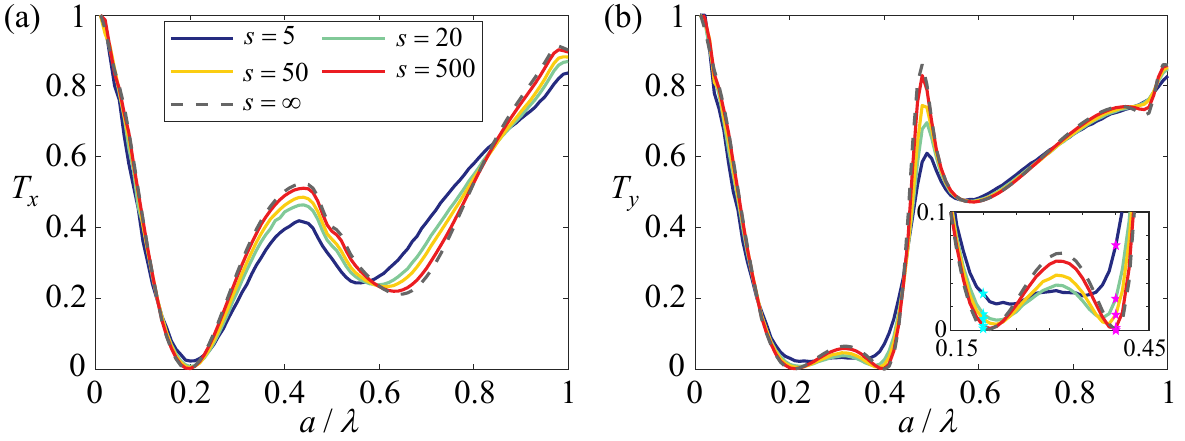}
	\caption{(a)(b) Transmissions $T_x$ and $T_y$ as functions of the lattice constant $a/\lambda$. The detuning is fixed at $\delta/\gamma = 0.5$. Different curve colors denote different values of the normalized trap depth $s$, as indicated in the legend of (a). The inset in (b) shows a magnified view, where the cyan and magenta stars mark the value at $a/\lambda = 0.2$ and $a/\lambda = 0.4$, respectively. Calculations assume an $N = 26 \times 26$ atom array and a diagonally polarized incident Gaussian beam with waist $w_0 = 0.3\sqrt{N}a$. Each line is an average of one hundred realizations.} 
	\label{TxTy_s}
\end{figure} 

In the first case, we evaluate the influence of positional uncertainty on the transmissions.
As shown in Fig.~\ref{TxTy_s}, we consider the representative detuning $\delta/\gamma=0.5$ and examine the behavior of the transmission zeros of $T_x$ and $T_y$.
In the $s\to\infty$ limit, atoms are perfectly localized, reproducing the ideal curves in Fig.~3(a,b) of the main text.
As the trap depth decreases, the spatial spread of the atomic wave packets increases, enhancing positional disorder in the array.
Consequently, the transmission zeros gradually shift away from zero.
Nevertheless, even for relatively shallow traps in our calculations ($s=5$), the transmission $T_y$ remains at the level of $\sim7.2\%$ at $a/\lambda=0.4$ and $\sim3.2\%$ at $a/\lambda=0.2$.
The corresponding transmission $T_x$ remains $\sim 2.4\%$ at $a/\lambda=0.2$.
These results indicate that the cooperative optical response of the dual-species array is robust against experimentally relevant positional disorder.

\begin{figure}[h!]
	\centering
	\includegraphics[width=0.75\linewidth]{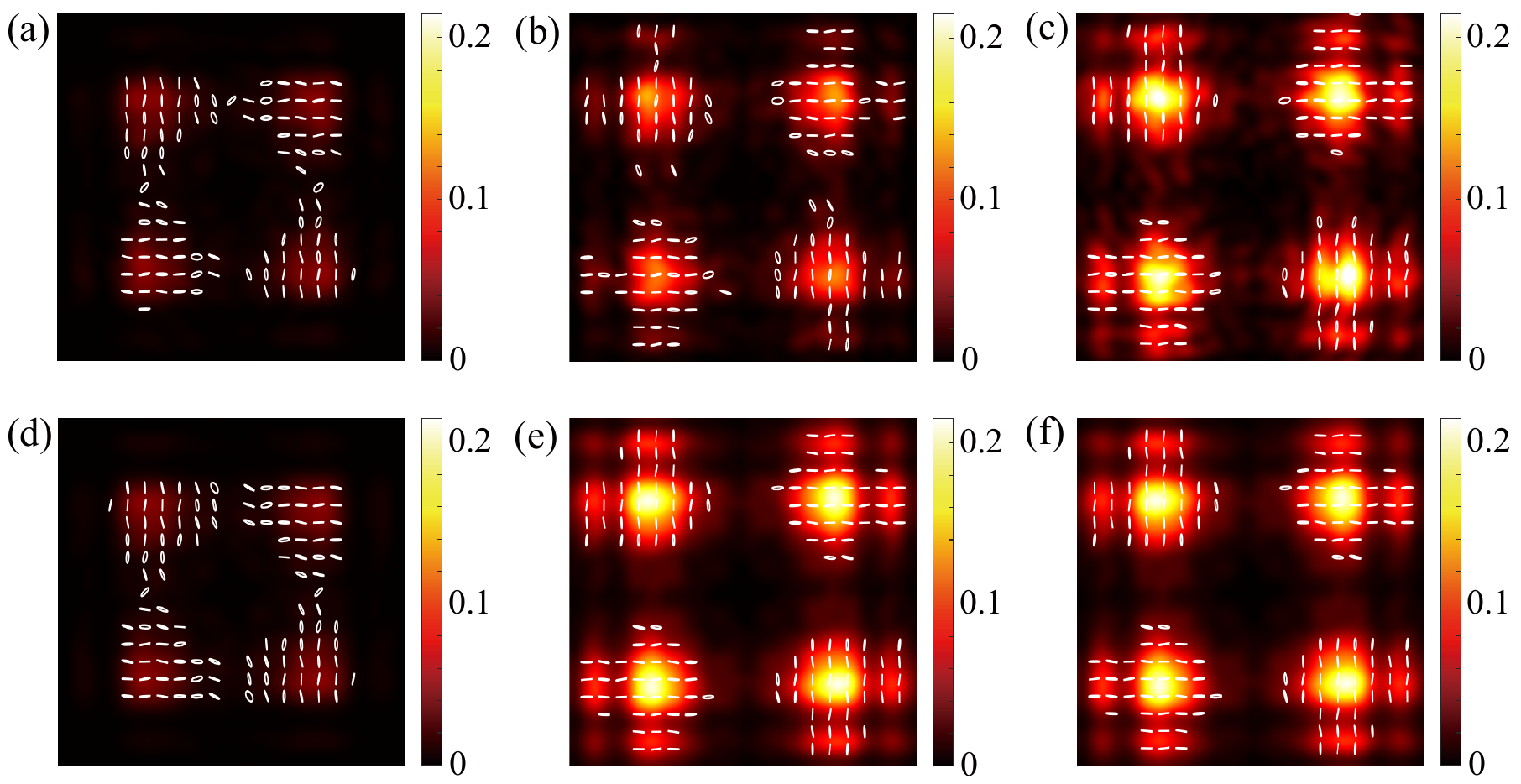}
	\caption{Total light intensity and polarization distribution at $z/\lambda = 14$ for atom arrays with lattice constant $a/\lambda = 0.4$ and different array sizes. Panels (a)–(c) correspond to the normalized trap depth $s = 500$, while (d)–(f) correspond to the idealized limit $s \to \infty$. For each configuration the array sizes are (a,d) $N = 51 \times 51$, (b,e) $N = 71 \times 71$, and (c,f) $N = 91 \times 91$. An isolation region of 11 lattice sites is included. Calculations assume a diagonally polarized incident beam with waist $w_0 = 0.3 \sqrt{N} a$ and detuning $\delta / \gamma = 0.5$. The color bar represents the normalized optical intensity $I = (|E_{x}|^{2} + |E_{y}|^{2})/|E_{0}|^{2}$, where $E_0$ is the peak amplitude of the incident Gaussian beam. White lines and ellipses indicate the polarization distribution. The figures span spatial ranges $x/\lambda, y/\lambda \in [-10, 10]$.}
	\label{polarization_32}
\end{figure} 

In the second case, we also examine array-size effects and the influence of positional uncertainty in the polarization-selective modulator configuration.
Fig.~\ref{polarization_32}(d–f) shows the total transmitted field intensity and polarization distribution for ideal $51\times51$, $71\times71$, and $91\times91$ arrays.
It shows that larger arrays, such as $71\times71$ and $91\times91$, yield nearly identical modulation, while the smaller $51\times51$ array exhibits weaker intensity and reduced polarization contrast.
Then, once positional uncertainty are included ($s=500$), Fig.~\ref{polarization_32}(a–c) reveals deviations in the local transmitted field intensity, but the overall modulation persists, with the $91\times91$ array most closely reproducing the ideal pattern.
These results indicate that large arrays are robust against positional uncertainty.

\section{Light shift of Yb atoms in a dipole trap}

For Yb atoms trapped in an optical dipole potential, the AC Stark effect induced by the oscillating electric field of the trapping light leads to energy shifts of the atomic levels. 
As a result, the transition frequency between a lower and an upper level is shifted, which can be expressed as~\cite{Brown2017}
\begin{equation}
	\Delta \nu = - \Delta\alpha U - \Delta\beta U^2,
\end{equation}
where $U$ denotes the trapping depth. 
Here $\Delta \alpha$ and $\Delta \beta$ represent the differences in the E1 polarizability and hyperpolarizability, respectively, between the lower and upper levels.
Since the trapping depth $U$ is proportional to the light intensity~\cite{Ushijima2018}, the frequency shift can be rewritten as
\begin{equation}
	\Delta \nu = - \Delta\alpha' \mathcal{I} - \Delta\beta' \mathcal{I}^2,
\end{equation}
where $\mathcal{I}$ denotes the light intensity.

Based on previous experimental measurements of the $1389\mathrm{nm}$ transition ($6s6p~{}^3P_0 \rightarrow 5d6s~{}^3D_1$) of $^{171}$Yb atoms in optical lattices with a typical magic wavelength of $759\mathrm{nm}$~\cite{Ai2023,Ai2024}, we can estimate the coefficients $\Delta\alpha'$ and $\Delta\beta'$. For optical lattices, we take $\mathcal{I} = 4\times 2P/(\pi w^2)$, where $P$ is the incident laser power and $w$ is the waist radius. 
From these parameters~\cite{Ai2023,Ai2024}, we obtain
\begin{equation}
	\Delta\alpha' = 1.04\times 10^{-9} \mathrm{MHz\cdot m^2/W}, \qquad \Delta\beta' = -2.3\times 10^{-19} \mathrm{MHz\cdot m^4/W^2}. 
\end{equation}
Next, we consider $^{171}$Yb atoms trapped in optical tweezers with wavelength $759\mathrm{nm}$. For a typical waist radius of $0.5\mathrm{\text\textmu m}$ and single-site powers of $4\mathrm{mW}$, $5\mathrm{mW}$, $6\mathrm{mW}$, and $8\mathrm{mW}$, the corresponding frequency shifts are approximately $13.3\mathrm{MHz}$, $24\mathrm{MHz}$, $37.8\mathrm{MHz}$, and $74.3\mathrm{MHz}$, respectively.

The frequency difference between the $1389\mathrm{nm}$ transitions of $^{171}\mathrm{Yb}$ and $^{173}\mathrm{Yb}$ atoms is about $27\mathrm{MHz}$ ($\approx 4.55\gamma$). 
To achieve $\delta < \gamma$, as required in the main text, this difference must be reduced below $2\gamma \approx 12\mathrm{MHz}$. 
This can be realized by trapping the two atomic species with different trap depths, forming a superlattice that reproduces the desired atomic-array pattern while generating distinct light shifts for the two species. 
As discussed above, the achievable shifts can reach several tens of MHz, enabling effective tuning of the relative detuning to $\delta < \gamma$. 

In practice, the transition frequencies of the $1389\mathrm{nm}$ line for $^{171}\mathrm{Yb}$ and $^{173}\mathrm{Yb}$ atoms under different trap depths would need to be carefully calibrated to determine the frequency-shift–versus–light-intensity relations and to identify the appropriate experimental parameters.

% Bibliography
\bibliography{references_1}